\documentclass[useAMS,usenatbib,usegraphicx]{mn2e}
\usepackage{times}
\usepackage{cite}

\usepackage{graphicx}
\usepackage{epsfig}
\usepackage{amssymb}
\usepackage{natbib}
\usepackage{journals}
\usepackage[latin1]{inputenc}
\newcommand{\Sauron}{\texttt{SAURON}}
\newcommand{\Muse}{\texttt{MUSE}}
\newcommand{\M}{M\,87}

\newcommand{\atlas}{{ATLAS$^{\rm 3D}$}}
\newcommand{\kms} {$\mbox{km s}^{-1}$}

\newcommand{\Msun} {$\mbox{M}_{\sun}$}

\def\spose#1{\hbox to 0pt{#1\hss}}
\def\lta{\mathrel{\spose{\lower 3pt\hbox{$\sim$}}
    \raise 2.0pt\hbox{$<$}}}
\def\gta{\mathrel{\spose{\lower 3pt\hbox{$\sim$}}
    \raise 2.0pt\hbox{$>$}}}
\newdimen\hssize
\newdimen\hdsize
\title[A kinematically distinct core in \M]
{A kinematically distinct core and minor-axis rotation: the \Muse\ perspective on \M}
\author
[Eric Emsellem et al.]{\parbox{\textwidth}{Eric Emsellem,$^{1,2}$\thanks{E-mail: eric.emsellem@eso.org \texttt{}}
Davor Krajnovi\'{c},$^{3}$ Marc Sarzi$^{4}$\vspace{0.4cm}}\\
\parbox{\textwidth}{$^{1}$European Southern Observatory, Karl-Schwarzschild-Str. 2, 85748 Garching, Germany\\
$^{2}$Universit\'e Lyon 1, Observatoire de Lyon, Centre de Recherche Astrophysique de Lyon \\ \hspace*{0.5cm} and Ecole Normale Sup\'erieure de Lyon, 9 avenue Charles Andr\'e, F-69230 Saint-Genis Laval, France\\
$^{3}$Leibniz-Institute f\"ur Astrophysics Potsdam (AIP), An der Sternwarte 16, D-14482 Potsdam, Germany \\
$^{4}$Centre for Astrophysics Research, University of Hertfordshire, Hatfield, Herts AL1 9AB, UK
} }
\begin{document}
\maketitle
%
%
\begin{abstract}
    We present evidence for the presence of a low-amplitude kinematically distinct component in the giant early-type galaxy  \M, via datasets obtained with the \Sauron\ and \Muse\ integral-field spectroscopic units.  The \Muse\ velocity field reveals a strong twist of $\sim$\,140\degr\ within the central 30\arcsec\ connecting outwards such a kinematically distinct core to a prolate-like rotation around the large-scale photometric major-axis of the galaxy. The existence of these kinematic features within the apparently round central regions of \M\ implies a non-axisymmetric and complex shape for this galaxy, which could be further constrained using the presented kinematics. The associated orbital structure should be interpreted together with other tracers of the gravitational potential probed at larger scales (e.g., Globular Clusters, Ultra Compact Dwarfs, Planetary Nebulae): it would offer an insight in the assembly history of one of the brightest galaxies in the Virgo Cluster. These data also demonstrate the potential of the \Muse\ spectrograph to uncover low-amplitude spectral signatures.
\end{abstract}

\section{Introduction\label{sec:intro}}

\M\ is the second brightest galaxy of the Virgo cluster, after M\,49, M\,86 being the third most luminous
member. Galaxies in the Virgo cluster mainly concentrate around these three giant early-type galaxies with 
the largest sub-structure centred on \M\ (Virgo~A), the other two (Virgo~B and C) respectively around M\,49 and
M\,60. \M\ itself is an extreme system, often classified as a cD due to its diffuse stellar envelope and halo
\citep{Weil1997, Mihos2005}, and
with a mass in stars of about $10^{12}$~\Msun\ \citep[e.g.][]{2011ApJS..197...33S,Murphy2011,2012ApJ...748....2D}. \M\ lies close to the centre of mass of the Virgo cluster as
traced by the X-ray emission \citep{Bohringer1994, Churazov2008}. It hosts an active supermassive black hole of a few billion solar masses
\citep{Sargent1978,Macchetto1997,Gebhardt2011,Walsh2013}, 
responsible for the triggering of a well-known radio jet \citep{Biretta1991}, and is
surrounded by more than 10,000 globular clusters \citep[GCs,][and Durell et al., submitted]{Peng2008}.
Its relative proximity \citep[with a mean distance of 16.5~Mpc for the Virgo cluster, and a distance of
16.7~Mpc for \M,][]{Mei2007} 
made it a target of choice for numerous studies.

Galaxy merging is assumed to play a prominent role in the hierarchical formation and evolution of such massive galaxies in clusters. 
A picture including a two-phase formation scenario recently emerged, with a first early gas-rich and violent merging stage, and a second
less dramatic assembly stage when stars are being accreted as stellar systems merge with the existing central
object \citep[see e.g.][]{DeLucia2007,Khochfar2009,Oser2010}. This history can in principle be traced in today's stellar population mix as well as in the
stellar kinematics of the galaxy. In that context, sub-structures such as kinematically decoupled cores
\citep{Franx1988,Jedrzejewski1988,Bender1988}, or kinematically distinct
components (KDC) as quantitatively defined in \citet{Krajnovic2011}, 
and triaxial figures have been called as important signatures of the violent merging past. 
For instance, major galaxy mergers are expected to  produce specific orbital stellar structures, and leave an imprint on the
observed velocity moments \citep{Balcells1998, Hernquist1991, Jesseit2007,  Hoffman2010, Bois2011}. KDCs have been observed
in many early-type galaxies \citep{Davies2001, Emsellem2004, Krajnovic2011}, 
and the modelling of the stellar kinematics led to interesting constraints 
on their intrinsic structures \citep{Statler2004, vandenBosch2008}.

Many past studies have reported observed stellar kinematics 
of \M, via long-slit spectroscopy \citep[e.g.][]{Sargent1978,Davies1988,Jarvis1991,vanderMarel1994a,
Bender1994} and integral-field units \citep[IFUs, see e.g.][]{Emsellem2004, Gebhardt2011, Murphy2011}. 
Most of these studies emphasise the lack of mean stellar rotation in
\M, consistent with the very low ellipticity within the central 30\arcsec. 
The isophotes have an increasing flattening towards the outer parts reaching
an ellipticity $\epsilon \sim 0.4$ at radial distances of a few hundreds of arcseconds \citep{Carter1978,Liu2005}.
\citet{Davies1988} tentatively suggested a minor-axis rotation (around a PA of 80\degr) at an amplitude
of about 20~\kms, although this result was cautiously flagged as marginally significant considering the modest
signal-to-noise ratio of the dataset. However, \citet{Jarvis1991} reported mean stellar rotation along the
minor-axis emphasising that "\M\ is not an oblate rotator". The stellar halo of M\,87 was also probed
with the VIRUS-P IFU, and \citet{Murphy2011} seem to detect mild rotation at large radii 
although it is not clear from their Figure~4 alone along which axis. More recently, 
a clear rotation pattern with an amplitude of $\sim 20$~\kms\ was revealed by the multi-slit approach of
\citet{Arnold2013}: the stellar rotation axis at a radius of 80\arcsec\ 
is not too far from the photometric major-axis of the galaxy. \M\ has been modelled
either as a spherical galaxy \citep[e.g.][]{vanderMarel1994a, vanderMarel1994, Zhu2014} 
or as an oblate system \citep[e.g.][]{Gebhardt2009}

The central region of \M\ has been recently observed with the \Muse\ spectrograph mounted at the ESO Very Large
Telescope (Paranal, Chile). In this letter, we thus revisit the stellar kinematics of \M\ within the central 30\arcsec\ ($\sim 2.5$~kpc) 
using integral-field spectroscopy, and reveal the presence of a KDC in the central region of \M\ using both \Sauron\ and \Muse.
We also clearly detect the signature of a prolate-like (minor-axis) rotation at a radius of about 30\arcsec.
In Section~\ref{sec:data}, we present the datasets, and the associated data reduction and analyses we have
performed. In Section~\ref{sec:results}, we focus on the resulting stellar kinematic maps and the revealed
kinematic structures. We briefly discuss these results in Section~\ref{sec:conc} and conclude.

\section{Data reduction and analysis}
\label{sec:data}

We describe here the \Sauron\ and \Muse\ datasets
that we are using to derive the stellar kinematics. We
focus on the extraction of the stellar kinematics and concentrate
in particular on the stellar velocity maps. A more complete
description of the \Muse\ observations for \M, the data reduction and
data quality assessment will be presented elsewhere (Sarzi
et al., in prep.).

\subsection{\Sauron\ Data}

\M\ has been observed with the \Sauron\ spectrograph \citep{Bacon2001}, an IFU installed at the
William Herschel Telescope in the Canary Islands. The dataset was obtained in the course of the \Sauron\
project \citep{deZeeuw2002} during Run~\#5 \citep[for details, see][]{Emsellem2004}, and corresponds to the merging of 3 \Sauron\
fields. Final spaxels before adaptive binning is performed are squared with 0\farcs8 on the side, the spectral
sampling and instrumental dispersion ($\sigma_{\rm spec}$) being 60 and 98~\kms, respectively.
We re-analysed the \Sauron\ \M\ dataset proceeding in the same fashion as described in \citet{Emsellem2004} 
and \citet{Cappellari2011}, but this time trying to probe low amplitude velocity structures within the central
region \citep[see][]{Krajnovic2011,Emsellem2011}. 
We therefore use a significantly higher threshold for the Voronoi binning scheme \citep{Cappellari2003}
with a minimum signal-to-noise ratio (SNR) of 300. Note that this value is only set as a 
relative goal as systematics do not ensure that the actual SNR increases steadily 
as spaxels are accreted.

\subsection{\Muse\ Data}

\M\ has recently been observed in the context of Programme
60.A-9312 (PI Sarzi) of the first \Muse\ Science Verification run conducted with
the Very Large Telescope at Paranal. \Muse\ delivers an impressive set
of 90,000 spectra covering most of the optical domain from about 4800
to 9000\,\AA\, with a spectral instrumental dispersion of about 60\,\kms\ at 5500\,\AA, 
and a field of view of nearly 1\arcmin\ square sampled in
rectangular $0\farcs2\times0\farcs2$ spaxels. The central regions of
\M\ were observed for 1h, and after reducing the data with version 0.18.1 of the \Muse\ pipeline (to be described
in Weilbacher et al., in prep.) these data delivered a mean SNR of 24 per spaxel at 5100\,\AA\ (for details on
the observations, data reduction and quality assessment, see Sarzi et
al. in prep). For consistency with our re-analysis of the
\Sauron\ data, we then spatially binned this datacube to 
achieve a similar target SNR of 300 using the Voronoi binning scheme.

\subsection{Stellar kinematics}

\begin{figure}
\centering
\epsfig{file=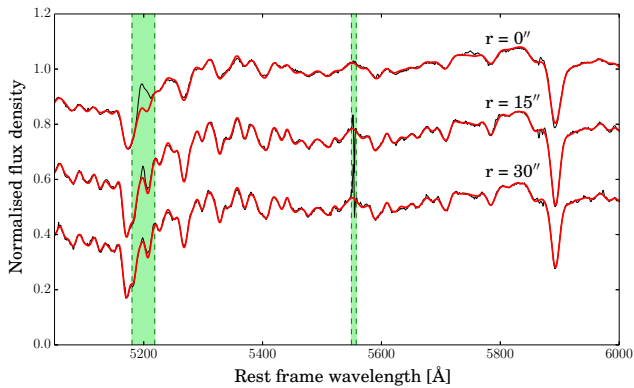, width=\columnwidth}
\caption{Three \Muse\ spectra (black lines) at three different radii (in the rest frame wavelength), 
with their corresponding pPXF fit (red
lines). The two regions which were discarded from the fit are indicated by the filled green areas, the first
corresponding to the [{\sc N$\,$i}]$\lambda\lambda5198,5200$ doublet, the second to a residual sky emission line.}
\label{fig:MuseSpectra}
\end{figure}

To extract the stellar kinematics from both the \Sauron\ and
\Muse\ dataset we used the penalised pixel fitting algorithm (pPXF)
developed by \citet{Cappellari2004}. For the \Sauron\ data, we used
the same mixed library of stellar and population model templates and
the same pPXF setup as done in \citet{Cappellari2011} 
in the case of the entire \atlas\ early-type galaxy sample
\citep[see also][]{Emsellem2004}. For the \Muse\ data, we
applied pPXF to the 5050 -- 6000\,\AA\ wavelength interval while
excluding regions potentially affected by 
[{\sc N$\,$i}]$\lambda\lambda5198,5200$ emission, adopting a 
10$^{\rm th}$ order additive polynomial correction for the continuum and
using a set of pre-derived templates that are best suited to match the
stellar populations of \M.
The latter templates correspond to the best combination of
stars in the entire MILES library \citep{Sanchez-Blazquez2006,
Falcon-Barroso2011} that match, with pPXF, nine
high-SNR spectra obtained by co-adding all the MUSE spectra within the
central 3\arcsec\ and in eight 3\arcsec-wide annular and circular
apertures extending out to 27\arcsec\ (see Sarzi et al. in
prep).
Fig.~\ref{fig:MuseSpectra} illustrates the quality of the MUSE data and of our pPXF fit
for three Voronoi binned spectra across the central 1\arcmin\ of
\M. We estimated the systematic uncertainty in the \Muse\ mean velocity measurements 
for these bins to be less than 2\,\kms\ both from fitting the unresolved sky lines and from the 
high frequency noise locally derived from the map itself (Fig.~\ref{fig:Vfields}).

\section{The central velocity field of \M}
\label{sec:results}

\begin{figure*}
\centering
\epsfig{file=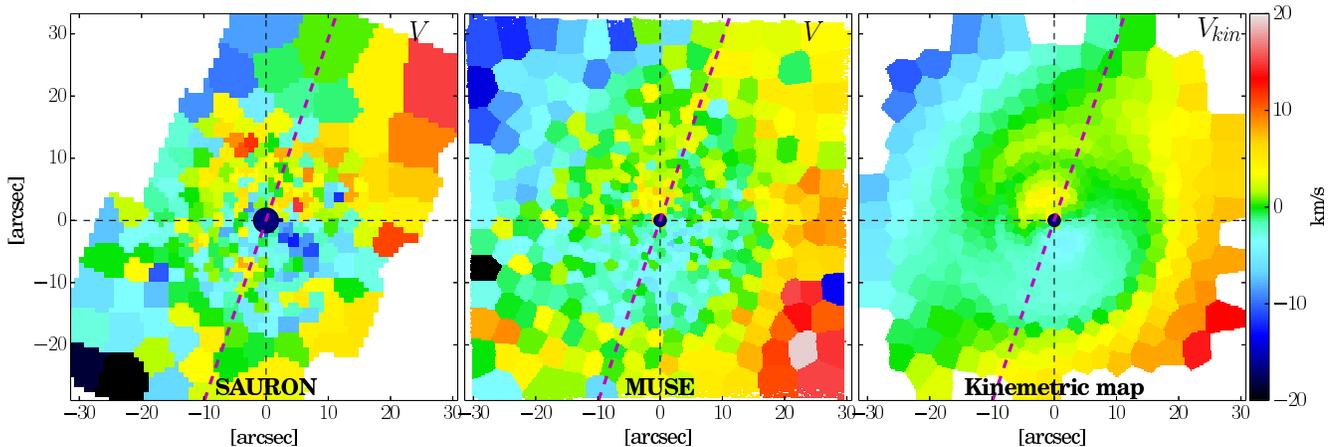, width=\hsize}
\caption{\Sauron\ (left panel), \Muse\ (middle panel) and reconstructed kinemetric (right panel) stellar velocity maps 
of the central region of \M. The three red crosses in the middle panel correspond to the positions
of the spectra shown in Fig.~\ref{fig:MuseSpectra}. The magenta dotted line marks the photometric major-axis of \M\ at large radii.
North is up, and East is left. The central 2\arcsec\ (left panel) or 1\arcsec\ (middle and right panels), 
where the extracted kinematics is significantly affected by the presence of very broad and strong emission lines, are masked. }
\label{fig:Vfields}
\end{figure*}

We first present the \Sauron\ stellar velocity of \M\ in the left panel of Fig.~\ref{fig:Vfields} where the Voronoi binning has been pushed up to reveal
low amplitude structures. Two odd point-symmetric regions emerge.
Firstly, the central 15\arcsec\ exhibit a rotation-like pattern with an amplitude of about $\pm 5$~\kms\ 
roughly aligned with the North-South direction, with the positive velocity in the North.
Secondly, a rotation pattern ($\pm 12$~\kms) is visible at the edge of the \Sauron\ field of view, but this time with an axis
roughly perpendicular to the central rotation pattern. Even though the formal errors are relatively low,
at a level of 4~\kms\ on average, the noise level is still high in this map
due to the combined modest spectral resolution, systematics and the difficulty to extract accurate information 
from the \Sauron\ short wavelength domain. This renders the detection of such low amplitude velocity structures 
tentative.

\begin{figure}
\centering
\epsfig{file=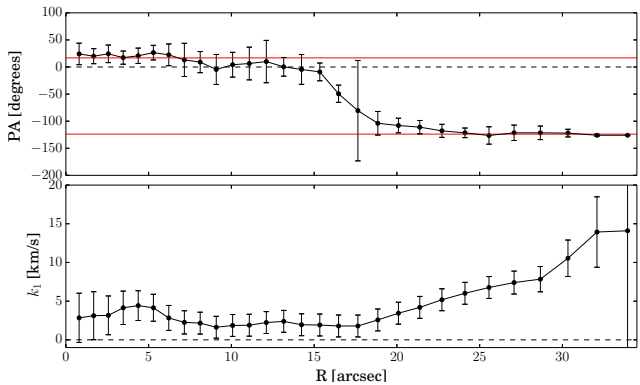, width=\columnwidth}
\caption{Position angle (top panel) and amplitude (bottom panel) from the extracted kinemetric profiles of the \Muse\ stellar velocity
field. The black dashed lines delineate the 0 and 360\degr\ limits, while the red dashed lines show the
average PA values within 13\arcsec\ and between 17 and 34\arcsec, respectively.}
\label{fig:rec_kinV}
\end{figure}

These features are, however, beautifully confirmed by the \Muse\ stellar velocity map (central panel of Fig.~\ref{fig:Vfields}).
The \M\ velocity field shows a kinematically distinct component within about 15\arcsec\, twisting outwards
connecting to a velocity pattern nearly symmetric around the photometric major-axis of the galaxy.
To quantify this further, we applied the kinemetry \citep{Krajnovic2006} to the \Muse\ mean
stellar velocity map, constraining the ellipticity of the fitted ellipse to be below 0.1, as to follow the
underlying photometry. The resulting profiles for the position angle and velocity amplitude are presented in
Fig.~\ref{fig:rec_kinV}, and the velocity field reconstructed using only up to the first order kinemetric harmonic is shown
in Fig.~\ref{fig:Vfields} (right panel).

The resulting kinematic position angle (PA) exhibits a clear and rather rapid 
change at a radius of $\sim 15$\arcsec\, where PA goes from a mean of 17\degr\ to about -124\degr\ (or 236\degr\ when measured East of North). This kinematic twist of about 140\degr\ has the same sign, but a larger amplitude than the photometric twist \citep[see e.g.][]{Ferrarese2006}, and is rather abrupt with a rapid transition region between 15 and 20\arcsec. The misalignments of the inner and the outer components is significant, i.e., the components are not at 180\degr\ misalignment and, hence, not strictly counter-rotating. This suggests a triaxial structure with, under the assumption that the density of the galaxy is stratified on similar ellipses, a viewing angle likely not along, but not far from, the intrinsic long axis \citep{Statler1991}. A more precise determination of the viewing angles requires the construction of detailed dynamical models \citep[e.g.][]{vandenBosch2009}, beyond the scope of this paper. 

\begin{figure}
\centering
\epsfig{file=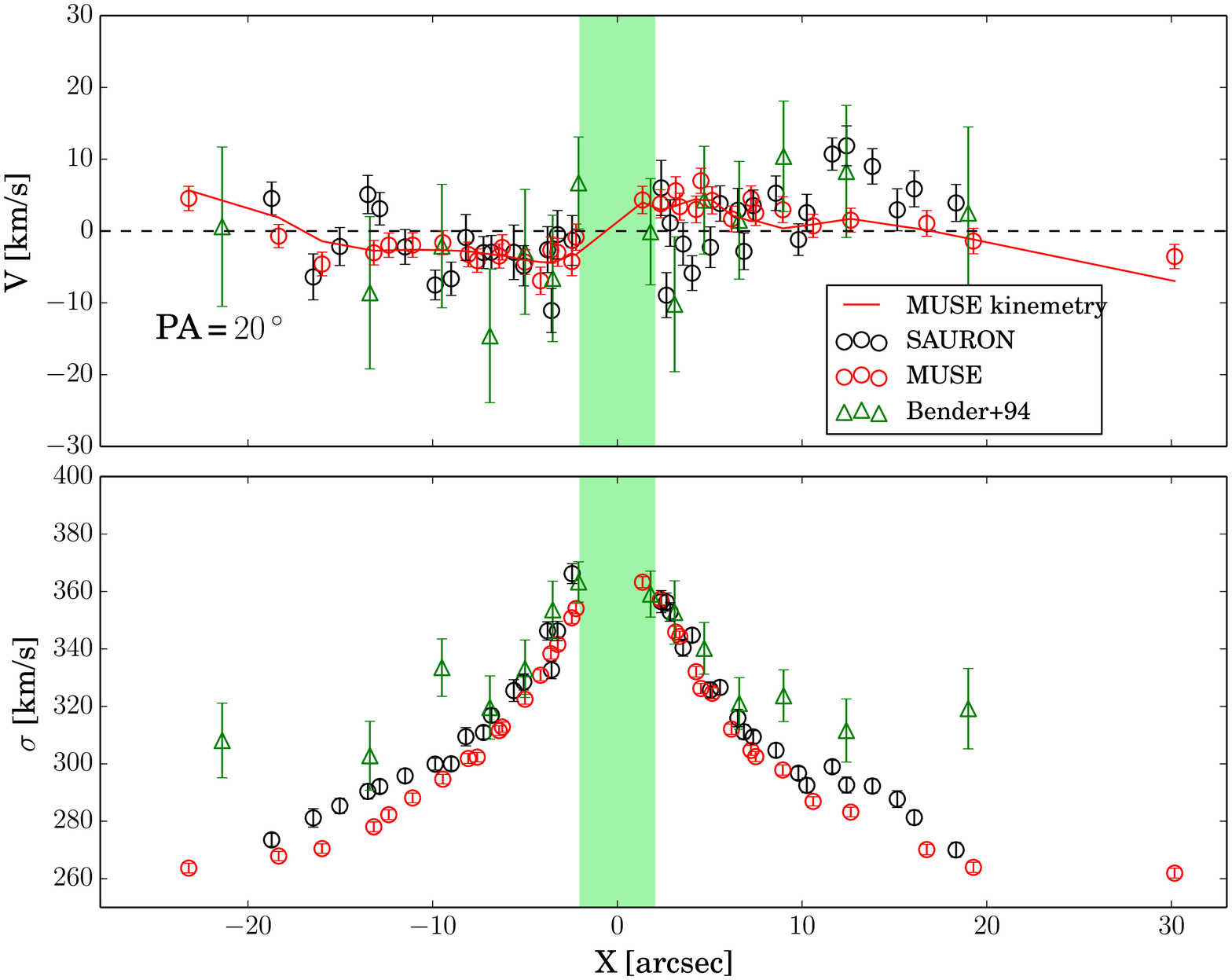, width=\columnwidth}
\caption{Cuts along a position angle PA$=20\degr$ of the \Muse\ (red circles) and \Sauron\ (black circles) 
stellar velocity $V$ (top panel) and velocity dispersion $\sigma$ (bottom panel) fields. The long-slit kinematics along the same PA 
from \citet{Bender1994} are overimposed
(green triangles). The red line shows the first harmonic kinemetric fit to the \Muse\ velocity field.
The green area shows the central 2\arcsec, where the \Sauron\ kinematics is deemed not reliable.
}
\label{fig:cut20}
\end{figure}

In the outer part, the axis of rotation is roughly aligned with the photometric minor-axis of \M, hinting for a prolate-like rotation. 
The stellar kinematics observed at the edge of the \Muse\ field of view connect well with the structure 
observed at larger radii by \citet{Arnold2013}. In Fig.~\ref{fig:cut20}, we also show that the \Muse\ stellar
kinematics is indeed consistent with the (noisier) \Sauron\ data, and with long-slit kinematics from
\citet{Bender1994}. The latter two include an odd symmetry in the radial velocity profile, but systematics
are on the high side. This contrasts with the low systematics in the \Muse\ dataset which is reflected in the 
average velocity uncertainty of 1.5~\kms, with most of the formal errors ranging from 1 to 2~\kms. Note that the
stellar velocity dispersion from \Sauron\ and \Muse\ are consistent with each others, while the long-slit
values from \citet{Bender1994} seem to overshoot in the outer part.

It is worth emphasising here that the present discovery of a central kinematically distinct 
component in the inner region of \M\ is an excellent illustration of the capabilities
of an IFU like \Muse\ which should therefore become an instrument of choice when 
exquisite two-dimensional information with low systematics are needed.

\section{Concluding remarks}
\label{sec:conc}

We have shown evidence for the existence of a distinct kinematic component in the central
15\arcsec of \M, using \Sauron\ and \Muse\ integral-field data. 
The rotation pattern of the core is of low amplitude ($\pm5$~\kms) and is 
offset by about 30\degr\ from the outer photometric major-axis. This connects via a 140\degr\ 
velocity twist to a minor-axis rotation in the outer part of the \Muse\ field. 
The classification of \M\ as a ``non-rotator'' \citep{Emsellem2004} should therefore be revised,
with \M\ joining the group of massive slow rotators with KDCs and outer minor-axis rotation such
as NGC\,4365 and NGC\,4406 \citep{Davies2001, Emsellem2004, Krajnovic2011}. 
This also questions the generic presence of KDCs in slow rotators \citep{Emsellem2011, Krajnovic2011}, while
already more than 60\% of all slow rotators in the \atlas\ sample have KDCs.

If the overall very low ellipticity of the isophotes in the central 30\arcsec\ is the result of a low 
inclination angle, this could imply a rather significant intrinsic mean stellar rotation in 
that region. However, it is worth noting that with a stellar velocity dispersion in 
excess of 250~\kms\ within 30\arcsec, the central dynamics of \M\ is very probably dominated by random motions. 
Still, the existence of the KDC and the measure of a velocity twist 
within its apparently round central region are key ingredients to
constrain the viewing angles and the shape of \M. It first triggers a qualitative
change of how we view the central region of this well-known massive galaxy: \M\ is often
considered as a spherical system while the present observations imply a complex non-axisymmetric structure with a radial
change in the orbital configuration. As emphasised by \citet{vandenBosch2009}, the detection of a radial
variation in the photometric position angle, and more importantly in the kinematic major-axis \citep[hence with
the zero velocity curve not being a straight line as in e.g., oblate systems, see][]{Statler1991}
is also a crucial ingredient for the recovery of the intrinsic shape: it fundamentally 
helps shrinking the parameter space to probe with such modelling. This is particularly true
for triaxial systems with KDCs where the uncertainty on the axis ratios is significantly reduced, even more so
when a strong twist and misalignment are present \citep{vandenBosch2009}. 
 
As shown by \citet{Hunter1992} for idealised models, the relative weights of the orbital families
varies roughly like the square of the ratio between the long-axis and short-axis.
From the observed KDC and kinematic twist, we can already assume a relative fractional increase of the importance of short-axis 
(resp. long-axis) tubes orbits in the inner (resp. outer) parts \citep[see e.g.][]{vandenBosch2008,Hoffman2010}. 
A more quantitative assessment of the global dynamics and orbital structure of \M\ 
would require detailed modelling of this new \Muse\ dataset, possibly complemented by a more extended \Muse\ mosaic, 
and adding e.g., the large-scale observations obtained by e.g., \citet{Arnold2013}.

Numerical simulations also suggest that a global trend exists between the importance 
of in-situ versus ex-situ star formation and e.g., the mean stellar age for
massive galaxies \citep{Naab2014}, and the orbital structure \citep{Jesseit2005}. 
The non-axisymmetry of \M\ both in its central part as probed by the misaligned KDC and in the outer part
with its minor-axis rotation already indicates a non-isotropic assembly.
Minor mergers are thus often called upon for as an important ingredient for the
assembly of massive galaxies in clusters \citep[see e.g.,][]{Burke2013}.
\citet{Naab2014} also emphasised the fact that the rare class of non-rotators 
has probably grown via the sole contribution of gas-poor minor mergers. 
The specific case of \M\ may tell us otherwise, with major mergers being an 
interesting process to consider here, given that a single gas-rich major event may be able 
to account for a significant part of the observed complexity \citep{Hoffman2010, Bois2011}.

The present kinematic observations, associated with 
a detailed study of the stellar populations in these regions should provide further constraints
on the merger history of \M. Other tracers such as Planetary Nebulae and Globular 
Clusters \citep{Kissler-Patig1998,Cote2001,Peng2008,Doherty2009,Longobardi2013} 
or Ultra Compact Dwarfs (UCDs) could help nailing down its evolutionary path.
An illustration of this comes from the recent claim by Zhang et al. (submitted) 
that the rotation axis of Ultra Compact Dwarfs (UCDs) orbiting \M\ is roughly orthogonal to that of the blue GCs. 
This may be for \M\ the potential missing link between the prolate-like rotation
of its stellar component emphasised in the present paper and the UCDs, while the blue GCs would follow
a more natural rotation pattern around the outer photometric minor-axis of the galaxy.
Whether UCDs are predominantly the remnants of harassed dwarfs, or the massive tail of stellar clusters formed during 
gas-rich merger events, a combined dynamical analysis of UCDs, GCs and stars in \M\, in comparison with
hydrodynamical simulations would bring our understanding of these structures one step further.
The ultimate goal would obviously be to reconcile all such signatures of the phase-space complexity of 
\M\ \citep{Romanowsky2012,Murphy2014} with a global view of its assembly history.

\section*{Acknowledgements}
The authors would like to warmly thank the entire \Muse\ team, and its PI (Roland Bacon),
for their many years of efforts to deliver such a remarkable instrument 
(hardware, operation software, advanced data reduction pipeline).
We would like to thank the \Muse\ Science Verification support team,
and Lodovico Coccato for comments on the data reduction. We also thank
Tim de Zeeuw for valuable input and comments on an early version of this manuscript.
EE dedicates this paper to Thilo. 

\bibliography{M87}{}
\bibliographystyle{mn2e}
\end{document}